\documentstyle[12pt,amstex,righttag,amsfonts]{article}

\newcommand{\C}{\Bbb{C}}
\newcommand{\R}{\Bbb{R}}
\newcommand{\Z}{\Bbb{Z}}
\newcommand{\M}{\Bbb{M}}
\newcommand{\bz}{\bar{z}}
\newcommand{\hN}{\hat{N}}

\newcommand{\A}{{\cal A}}
\renewcommand{\H}{{\cal H}}
\newcommand{\bea}{\begin{eqnarray}}
\newcommand{\ena}{\end{eqnarray}}
\newcommand{\vs}[1]{\vspace{#1 mm}}

\def\bbox{{\,\lower0.9pt\vbox{\hrule \hbox{\vrule height 0.2 cm
\hskip 0.2 cm \vrule height 0.2 cm}\hrule}\,}}
\newcommand{\pa}{\partial}
\newcommand{\nn}{\nonumber\\}
\newcommand{\lan}{\langle}
\newcommand{\ran}{\rangle} 
\makeatletter
 
 \@addtoreset{equation}{section}
\makeatother

\begin{document}


\topmargin 0pt
\oddsidemargin 0mm
\renewcommand{\thefootnote}{\fnsymbol{footnote}}


\begin{titlepage}

\setcounter{page}{0}
\begin{flushright}
KEK-TH 695\\
hep-th/0005199\\
\end{flushright}

\vs{15}
\begin{center}
{\Large\bf  Equivalence of Projections as Gauge Equivalence} \\
\vs{5}
{\Large\bf on Noncommutative Space}\\
\vs{20}
{\large
Kazuyuki \ Furuuchi}

\vs{10}
{\em Laboratory for Particle and Nuclear Physics,}\vs{1}\\
{\em High Energy Accelerator Research Organization (KEK), }\vs{1}\\
{\em Tsukuba, Ibaraki 305-0801, Japan} \vs{2}\\
{\em E-mail: furuuchi@@ccthmail.kek.jp}
\end{center}

\vs{15}
\centerline{{\large\bf{Abstract}}}
\vs{5}
\noindent Projections play
crucial roles in the ADHM construction 
on noncommutative $\R^4$.
In this article
a framework for the description of
equivalence relations between projections is proposed. 
We treat the equivalence of projections
as ``gauge equivalence'' on noncommutative space.
We find an interesting application of this framework 
to the study of $U(2)$ instanton on noncommutative $\R^4$:
A zero winding number configuration with a hole at the
origin is ``gauge equivalent'' to the noncommutative
analog of the BPST instanton.
Thus the ``gauge transformation'' in this case
can be understood as a noncommutative resolution
of the singular gauge transformation
in ordinary $\R^4$.

\end{titlepage}
\newpage

\renewcommand{\thefootnote}{\arabic{footnote}}
\setcounter{footnote}{0}

\section{Introduction}

The concept of 
smooth space-time manifold
should be modified at the Planck scale
due to the quantum fluctuations,
and  we except 
the  short scale structure of space-time
has noncommutative nature.
When the coordinates of the
space are noncommutative,
we except
the appearance of short scale 
cut off at the noncommutative scale.
For example,
instantons 
on noncommutative $\R^4$ 
constructed by the ADHM method \cite{NS}
never become singular \cite{mine},
due to the cut off in the size of instanton.\footnote{
This is the case when the 
noncommutativity of the coordinates has
self-dual part
(and instantons are anti-self-dual)\cite{SW}.} 
Although the noncommutativity 
in this case is quite simple,
the construction  reveals
deep insights in the nature of gauge theory 
on noncommutative space.
Indeed,
the precise mechanism that leads to the
absence of singularity
is quite nontrivial.
In order to construct instantons on 
noncommutative $\R^4$,
one needs to project out
some states in Hilbert space,
where the Hilbert space is introduced
to represent the algebra
of noncommutative $\R^4$.
Since noncommutative $\R^4$
is defined by the whole Hilbert space
and projection 
removes some of the states
in this Hilbert space,
projection can be interpreted 
as a change of topology of the
base manifold.
More precisely,
projection removes some points 
from $\R^4$ and creates holes.
Hence instantons
on noncommutative $\R^4$
indicates the necessity for
the unified description of 
gauge fields and geometry \cite{mine}\cite{BN}.

In this article a framework for the description of
equivalence relations between projections is proposed.
We treat the equivalence of projections
as a kind of gauge equivalence.
Hence the formalism of this framework is 
similar to the gauge theory. 
However since the
projection contains information
of the Hilbert space which represents 
noncommutative $\R^4$,
the transformation between
equivalent projections may be regarded
as a noncommutative 
analog of coordinate transformation.
Therefore this is a 
possible framework for the unified description of
gauge fields and geometry.
We find an
interesting application of this framework 
to the study of
$U(2)$ instanton on noncommutative $\R^4$.
\section{Equivalence of Projections as Gauge Equivalence on %
Noncommutative Space}

In this section we explain the notion of the
equivalence of projections in 
a concrete example,
the gauge theory on
noncommutative $\R^4$.
However it is obvious that 
following arguments can be extended to
gauge theory on
more general noncommutative space.

\subsection*{Reviews on Gauge Theory on Noncommutative $\R^4$}

The noncommutative
$\R^4$ we shall consider
is described by 
an algebra
generated by the noncommutative coordinates
$x^{\mu} \, \,  (\mu  = 1 ,\cdots , 4) $
which satisfy  the following commutation relations:
\bea
 \label{noncomx}
[ x^{\mu} ,  x^{\nu}] = i \theta^{\mu\nu} , 
\ena
where $\theta^{\mu\nu}$ is real 
and constant.
In this article we consider the case  
where the $\theta^{\mu\nu}$ is self-dual,
and set
\bea
 \label{theta}
\theta^{12} = \theta^{34} = \frac{\zeta}{4}  \, , \, 
\quad \zeta > 0
\quad \mbox{(others: zero)},
\ena
for simplicity.
Next we introduce the complex noncommutative
coordinates by 
\bea
z_1 = x_2 + i x_1  , \quad z_2 = x_4 + i x_3 \,  .
\ena
Their commutation relations become
\bea
 \label{noncom}
 [z_1 , \bar{z}_1] 
=[z_2 , \bar{z}_2] 
= - \frac{\zeta}{2}\,  \quad 
\mbox{(others: zero)}.
\ena
We start with 
the algebra End ${\cal H}$ 
of operators acting in the
Hilbert space\\
${\cal H} = \sum_{(n_1,n_2) 
\in \Z_{\geq 0}^2 }
\C \left| n_1 , n_2    \right\ran$, 
where 
$z$ and  $\bar{z}$ are represented
as creation and annihilation operators:
\bea
\sqrt{\frac{2}{\zeta} }
z_1 \left| n_1  , n_2    \right\ran
&=& \sqrt{n_1 +1} \left| n_1 + 1 , n_2    \right\ran , \quad
\sqrt{\frac{2}{\zeta} }
\bar{z}_1 \left| n_1  , n_2    \right\ran
= \sqrt{n_1 } \left| n_1 -1 , n_2    \right\ran , \nn
\sqrt{\frac{2}{\zeta} }
z_2 \left| n_1  , n_2    \right\ran
&=& \sqrt{n_2 +1} \left| n_1 , n_2 +1    \right\ran , \quad
\sqrt{\frac{2}{\zeta} }
\bar{z}_2 \left| n_1  , n_2    \right\ran
= \sqrt{n_2} \left| n_1 , n_2 -1    \right\ran .
\ena
The commutation relations in (\ref{noncomx})
have  automorphisms of the form
$ x^{\mu} \mapsto x^{\mu} + c^{\mu}$,
where $c^{\mu}$ is a commuting real number.
These automorphisms are
generated by unitary operator $U_c$:
\bea
U_c :=
\exp
[c^{\mu} \hat{\pa}_{\mu}] ,
\ena
where we have introduced
{\bf derivative operator}
$\hat{\pa}_{\mu}$ by
\bea
 \label{deri}
\hat{\pa}_{\mu} := i B_{\mu\nu} x^{\nu}.
\ena
Here $B_{\mu\nu}$ is a inverse matrix of $\theta^{\mu\nu}$.
$\hat{\pa}_{\mu}$ satisfies following
commutation relations:
\bea
 \label{comdel}
[\hat{\pa}_{\mu}, x^{\nu}] = \delta_{\mu}^{\nu}, \quad
[\hat{\pa}_{\mu}, \hat{\pa}_{\nu}]
= i B_{\mu\nu}.
\ena
One can check the following equation:
\bea
U_c\, x^{\mu}\, U_c^{\dagger} =\,  x^{\mu} + c^{\mu}.
\ena
We define {\bf derivative of
operators} $\hat{O} \in \mbox{End}\, {\cal H}$ by
\bea
\label{del}
\pa_{\mu} \hat{O} :=
\lim_{\delta c^{\mu} \rightarrow 0}
\frac{1}{\delta c^{\mu} } 
\left(
U_{\delta c^{\mu}}
\hat{O} U_{\delta c^{\mu}}^{\dagger} - \hat{O}
\right)=
[\hat{\pa}_{\mu}, \hat{O} ].
\ena
The action of two derivatives commutes:
\bea
\pa_\mu \pa_\nu \hat{O} - \pa_\nu \pa_\mu \hat{O}
=[\hat{\pa}_{\mu}, [\hat{\pa}_{\nu}, \hat{O} ]] 
- (\mu \leftrightarrow \nu)
= 0.
\ena
Operator $\hat{O}$ is called 
{\bf bounded operator} if
\bea
\forall \left| \phi  \right\ran \in 
\mbox{Dom}(\hat{O}) , \quad
|| \hat{O} \left| \phi  \right\ran|| \leq
C  || \left| \phi  \right\ran||,
\ena
for some constant $C>0 $, where
$\mbox{Dom}(\hat{O})$ is a domain of 
operator $\hat{O}$. 
The norm of bounded operators 
are defined by
\bea
|| \hat{O} || := 
\mbox{sup}\, \frac{||\, \hat{O} \left| \phi  \right\ran  || 
}{  \,|| \, \left| \phi  \right\ran  ||  }
, \, \,\phi \ne 0, \,  \left| \phi  \right\ran \in \mbox{Dom}(\hat{O}),
\ena
where sup means the supremum.
We call the operator
{\bf smooth} when the derivative
of the operator is a
bounded operator.
We shall consider
the algebra of smooth bounded operators
and denote this algebra by ${\cal A}$.

The  $U(n)$ gauge field
on noncommutative
$\R^4 $ is defined 
as follows.
First
we consider 
$n$-dimensional vector space
${\cal A}^n := \C^n \otimes {\cal A}$.
The elements of ${\cal A}^n$ can be thought of
as $n$-dimensional vectors with their
entries in ${\cal A}$.
Let us consider the 
unitary action 
on the element of ${\cal A}^n$:
\bea
\label{Ue}
\phi \rightarrow U\phi.
\ena
Here $U \in \M_n(\A)$ 
($\M_n(\A)$ denotes the
algebra of $n\times n$ matrices with
their entries in ${\cal A}$) and 
satisfying $UU^{\dagger}=U^{\dagger}U
=\mbox{Id}_{\M_n(\A)}$,
where $\mbox{Id}_{\M_n(\A)}$ is the identity
operator in $\M_n(\A)$.
In general $U$ depends on $z$ and $\bz$, and hence
we regard this unitary transformation as
gauge transformation.
We define the action of exterior derivative 
$d$ by
\bea
da := (\pa_{\mu} a )\,dx^{\mu}, \quad 
a \in {\cal A}. 
\ena
We define the covariant derivative of $\phi \in {\cal A}^n$ 
as a derivative which 
transforms covariantly
under the 
gauge transformation (\ref{Ue}), i.e.
\bea
\label{UDe}
 D\phi \rightarrow  UD\phi ,\qquad
D = d+A.
\ena
Here the $U(n)$ gauge field 
$A$ is  
introduced to ensure the covariance. 
$A$ is a matrix valued one-form:
$A = A_{\mu}dx^{\mu}$ and
$A_{\mu} \in \M_n({\cal A})$ 
is anti-Hermitian.
$dx^{\mu}$ commute with $x^{\mu}$
and anti-commute among themselves,
and hence
$d^2 a =0$ for $ a \in {\cal A}$.
From (\ref{Ue}) and (\ref{UDe}),
the covariant derivative 
transforms as
\bea
 D \rightarrow UDU^{\dagger} .
\ena
Hence the gauge field $A$ 
transforms as
\bea
A \rightarrow UAU^{\dagger}+UdU^{\dagger}.
\ena
The field strength
is defined by
\bea
F := D^2 = dA + A^2.
\ena
We can construct a gauge invariant
action $S$ as follows:
\bea
 \label{action}
S = -\frac{1}{g^2}
\left(2\pi\right)^2 \sqrt{ det \theta}\, \,
\mbox{Tr}\, \, 
F \wedge * F,
\ena
where Tr denotes the trace over $\H^n := \C^n \otimes \H$
and $*$ is the Hodge star.\footnote{In this paper
we only consider the case
where the metric 
on $\R^4$ is flat:
$g_{\mu\nu} = \delta_{\mu\nu}$.}
If we use the operator symbols and the star product, 
(\ref{action}) can be rewritten as\footnote{%
For the explicit form of
the map from operators
to operator symbols,
see for example \cite{NCIKKT}\cite{mine}.
}
\bea
S = -\frac{1}{4g^2} \int d^4x \, 
\mbox{tr}\, F_{\mu\nu} \star F^{\mu\nu}.
\ena
Here tr denotes the trace over the $U(n)$ gauge group.
In the above,
and throughout this article,
we use the same letters
for operators and corresponding operator symbols
for notational simplicity.

Next let us consider gauge theory with 
projection \cite{mine}.\footnote{
For the roles of projections
in noncommutative geometry, see for example
\cite{Con}\cite{Landi}.}
A projection 
$p$ is an Hermitian idempotent
element in $\M_n ({\cal A})$:
$p^{\dagger} = p$, $p^2=p$. 
We consider vector space 
$p {\cal A}^n :=
\{
\phi_p \in {\cal A}^n:
\phi_p = p \phi_p
\}$.
We can consider
a unitary action
on $p {\cal A}^n$ (which is unitary in the
restricted vector space $p {\cal A}^n$):
\bea
 \label{unip}
 \phi_p \rightarrow U \phi_p ,  
\quad 
U^{\dagger} U
=
U U^{\dagger}
= p . \,
\ena
We can  construct 
covariant derivative $D_{p}$ for
$p {\cal A}^n$ by
\bea
 \label{Pd1}
D_p = pd + A_p, \quad A_p = pA_p p.
\ena
We require $D_{p} \phi_p$ to transform covariantly
under the unitary transformation: 
\bea
D_p \phi_p \rightarrow 
U D_p \phi_p .
\ena
Then the covariant derivative
$D_p$ must transforms as
\bea
 \label{PUPDP}
D_p \rightarrow 
 U D_p 
U^{\dagger}.
\ena
For any $ {\phi'}_p \in p{\cal A}^n$, following
equation holds
\bea
U D_p
 U^{\dagger} {\phi'}_p
&=&  U (pd + A_p)  
U^{\dagger} {\phi'}_p
 =   U d 
(U^{\dagger} {\phi'}_p)
 + U A_p
 U^{\dagger} {\phi'}_p \nn
&=&  U
dU^{\dagger}{\phi'}_p
+  U 
(U^{\dagger} d{\phi'}_p )
    + U A_p
U^{\dagger} {\phi'}_p
    \quad (U=U p = p U)\nn
&=& p d{\phi'}_p + (U dU^{\dagger} + U A_p U^{\dagger}) {\phi'}_p.
\ena
Hence the gauge field
$A_p$ transforms as
\bea
A_p \rightarrow 
U A_p U^{\dagger}  
+ U(dU^{\dagger})p.
\ena
The field strength becomes
\bea
 \label{PF1}
F &:=& D_p^2  \nn
  &=& p (dA_p) p + A_p^2 + pdpdp .
\ena
Indeed, for arbitrary $\phi_p \in p{\cal A}^n$,
\bea
 \label{Fphi}
F \phi_p &=& (pd + A_p)(pd\phi_p + A_p \phi_p) \nn
   &=& pd(pd \phi_p) + pd(A_p\phi_p) + A_p pd\phi_p + A_p^2 \phi_p \nn
   &=& pd(pd \phi_p) + pdA_p \phi_p + A_p^2 \phi_p,
\ena
and since $\phi_p = p \phi_p$ and $p^2 = p$,
the term $pd(pd\phi_p ) $ in (\ref{Fphi}) becomes
\bea
pd(pd\phi) &=& pd(pd(p\phi_p )) \nn
        &=& pd(pdp\, \phi_p + pd\phi_p ) \nn
        &=& pdpdp\, \phi_p -pdpd\phi_p+pdpd\phi_p \nn
        &=& pdpdp\,  \phi_p.
\ena
We can construct 
action $S$ which is invariant under 
the unitary transformation (\ref{unip}):
\bea
 \label{pS}
S = -\frac{1}{g^2}
\left(
2\pi 
\right)^2
\sqrt{det\theta}\, \, 
\mbox{Tr}\,
F\wedge *F.
\ena

\subsection*{Equivalence of Projections \footnote{For 
detailed explanations on 
the equivalence of projections,
see \cite{Con}\cite{WO}. } }

However, 
there exists more larger class of transformations
under which 
the action (\ref{pS}) is invariant.
In this subsection
we will describe these transformations.
We start from the definition of the
equivalence of projections,
and then we treat the
equivalence relation  
as gauge equivalence.

Projections $p$ and $q$ in the algebra $\M_n(\A)$
are said to be equivalent,
or {\bf Murray-von Neumann equivalent}
when\footnote{%
Here we consider $\M_n(\A)$ as an example, but
Murray-von Neumann equivalence can be considered in any
$C^*$-algebra.
}
\bea
{}^\exists U \in \M_n(\A) ,\quad
p = U^{\dagger}U \quad  \mbox{and} \quad 
q = U U^{\dagger}, 
\ena
and denoted as $p \sim q$.
These operators satisfy following equations:
\bea
U = Up = q\, U ,\quad U^{\dagger} = p \, U^{\dagger} = U^{\dagger}q .
\ena
\bea
\mbox{Ker} \, U = \mbox{Id}_{\M_n(\A)} - U^{\dagger} U , \quad
\mbox{Ker} \, U^{\dagger} =  \mbox{Id}_{\M_n(\A)} - U U^{\dagger}.
\ena
\bea
 \label{up}
U^{\dagger} \H^{n} 
= U^{\dagger}U \H^{n} = p \H^{n}, \quad
U\H^{n} = UU^{\dagger} \H^{n} = q \H^{n}  .
\ena
By choosing 
orthonormal basis of $p \H^{n}$ and
$q \H^{n}$, it is easily seen that
\bea
 \label{pHqH}
p \sim q  \Leftrightarrow
\mbox{dim}\, p \H^{n} = \mbox{dim} \, q \H^{n}.
\ena
Note that
$p$ can be equivalent to the identity
if $p$ has infinite rank.
From (\ref{up}), $U$ can be regarded as a map from  $p \A^n$ to  $q \A^n$:
\bea
\phi_p \rightarrow \phi_q = U \phi_p, \quad
\phi_p \in p \A^n, \, \phi_q \in q\A^n   .
\ena
We require the covariant derivative of
$\phi_p$ is also mapped in the same form as  $\phi_p$:
\bea
 \label{pqG}
D_p \phi_p \rightarrow  U D_p \phi_p = D_q \phi_q ,
\ena
where $D_q = qd +A_q$ and $A_q = q A_q q$ is a 
transform of $A_p$.
This requirement determines
the transformation rule of gauge fields 
$A_p \rightarrow A_q$ uniquely:
\bea
D_q U \phi_p
&=& 
(qd + A_q) U \phi_p\nn
&=& 
UU^{\dagger} d (U \phi_p)  
+ U^{\dagger} A_q U \phi_p
=
U(pd + U^{\dagger} (d U ) + U^{\dagger} A_q U ) \phi_p.
\ena
Hence
\bea
 \label{Ap}
A_p = U^{\dagger}A_q U+ U^{\dagger} (d U ) p  .
\ena
Then, 
\begin{align}
U A_p U^{\dagger}
&=
A_q + UU^{\dagger} (d U ) pU^{\dagger} \nn
&=
A_q + q (d U ) U^{\dagger}q \nn
&=
A_q + q (d (U  U^{\dagger}) - U  d U^{\dagger}) q \nn
&=
A_q + q (dq  - U  d U^{\dagger}) q \nn
&=
A_q - U  (d U^{\dagger}) q.
\end{align}
Here we have used the basic 
identity for projections:
$q(dq)q = 0$.
Hence we obtain the reversal formula
of (\ref{Ap}) consistently:
\bea
 \label{Aq}
A_q = 
UA_pU^{\dagger} + U(d U^{\dagger}) q  .
\ena
The transformation rule 
(\ref{Aq})
is similar to
the usual gauge transformation,
and therefore we also call it
gauge transformation, or
{\bf Murray-von Neumann gauge transformation}
(MvN gauge transformation)
if we stress the difference from the 
usual gauge transformation on
noncommutative space.
MvN gauge transformation contains
the transformation proposed in \cite{Ho}
as a special case.\footnote{
However we regard that the rank of the projection
does not change under this transformation
as opposed to \cite{Ho}.
For example, $\mbox{Id}_{\H}$ and 
$\mbox{Id}_{\H} - |0,0\ran\lan 0,0 |$ can be
Murray-von Neumann equivalent since
both have infinite rank
(see eq.(\ref{pHqH})).
}
The transformation rule for the field strength
is obtained as
\bea
F_p  = D_p^2
\rightarrow
F_q &=&  D_q^2 \nn
&=& U D_p U^{\dagger} U D_p U^{\dagger}
     = U D_p p  D_p U^{\dagger}         \nn
&=& U D_p^2 U^{\dagger} = U F_p U^{\dagger}.
\ena
The important point is that
under the
MvN gauge transformation the action (\ref{pS})
is invariant.
This is because
\begin{align}
\mbox{Tr}\, \,  F_p \wedge * F_p
\rightarrow&\, \, 
\mbox{Tr}\, \,  F_q \wedge * F_q \nn
&=\,
\mbox{Tr}\, \,  U^{\dagger} F_p \wedge * F_p  U^{\dagger}\nn
&=\,
\mbox{Tr}\, \,   p F_p \wedge * F_p  p\nn
&=\,
\mbox{Tr}\, \,   F_p \wedge * F_p .
\end{align}
Here we have used eq.(\ref{up}).
The noncommutative
$\R^4$ is represented by operators End $\H$.
Hence one-to-one map between
Hilbert space may be regarded as a
noncommutative analog of coordinate
transformation.
The MvN gauge transformation $U$ can be regarded as a map from 
$p \H^{n}$ to $q \H^{n}$, and thus it 
can be understood as a mixture of gauge transformation 
and coordinate transformation on noncommutative $\R^4$.

\section{Application to Instanton on Noncommutative $\R^4$}

\subsection*{$U(2)$ One-Instanton Solution on Ordinary $\R^4$}

In order to illustrate the similarity and difference
between commutative and noncommutative case,
let us first construct the $U(2)$ one-instanton solution
by the ADHM method
in the case of ordinary commutative ${\R}^4$.
In this subsection, $z$ and $\bz$ represent ordinary 
commuting coordinates.

In order to construct instantons
by the ADHM method \cite{ADHMconst}, we start from the
following data:
\begin{enumerate}
 \item A pair of complex hermitian vector spaces
       $V =  \C^k $ and $W = \C^n $ . 
 \item The operators 
     $B_1 , B_2 \in Hom(V,V) ,
      I \in Hom(W,V) , J = Hom(V,W) $\ %
  satisfying the equations $\mu_{\R}=\mu_{\C}=0$, where
\begin{align}
 \label{ADHM}
\mu_{\R}
 &= [B_1 , B_1^{\dagger}] +  [B_2 , B_2^{\dagger}] 
       + II^{\dagger} - J^{\dagger} J, \\
\mu_{\C} &= [B_1 , B_2 ] + IJ.
\end{align}
\end{enumerate}
Next we define Dirac-like operator
${\cal D }_{z} :
V \oplus V \oplus W \rightarrow V \oplus V $ by
\begin{align}
 \label{Dz}
& {\cal D}_z = 
\left(
 \begin{array}{c}
  \tau_z \\
  \sigma_z^{\dagger }
 \end{array}
\right)  , \nn
& \tau_z =
(\, B_2 - z_2 ,\, B_1 - z_1 , \, I \, ) , \nn
& \sigma_z^{\dagger} =
(\, - (B_1^{\dagger} -\bar{z}_1) , \,
        B_2^{\dagger} - \bar{z}_2  , \, J^{\dagger} \, ) .
\end{align}
The equation $\mu_{\R}=\mu_{\C}=0$ 
is equivalent to the set of equations
\begin{align}
\label{key}
\tau_z \tau_z^{\dagger}
=\sigma_z^{\dagger} \sigma_z := \Box_z, \quad
\tau_z \sigma_z = 0 .
\end{align}
The second equation means
$\mbox{Im }\sigma_z \in \mbox{Ker}\, \tau_z $,
and therefore
dim $\mbox{Ker } \tau_z / \mbox{Im }\sigma_z 
= (2k+n-k)-k =n$.
Hence there are $n$ zero-eigenvalue-vectors
(we call them zero-modes for short) of 
${\cal D}_z$ :
\bea
 \label{czerom}
{\cal D}_z \Psi^{(a)} = 0 ,
\quad a= 1, \ldots , n.
\ena
We can choose orthonormal basis of the space
of the zero-modes:
\bea
 \label{ortho}
\Psi^{(a)\dagger } \Psi^{(b)} = 
\delta^{ab}.
\ena
There is a freedom in the choice of the basis:
\bea
\Psi \rightarrow \Psi U^{\dagger}, \quad
\Psi =
\left(
 \begin{array}{ccc}
            &        &  \\
 \Psi^{(1)} & \cdots & \Psi^{(n)} \\
            &        & 
 \end{array}
\right)  ,
\ena
where $U$ is an $n\times n$  unitary matrix.
$U$ may depends on $z$ and $\bz$, and
this change of basis will become $U(n)$ gauge symmetry
after we construct gauge fields from the zero-modes.
Anti-self-dual $U(n)$ gauge
field is constructed 
by the formula
\bea
\label{Amu}
A^{ab} 
= \Psi^{(a)\dagger } d \Psi^{(b)} ,
\ena
where $a$ and $b$ are indices of 
$U(n)$ gauge group.
There is an action of $U(k)$ that does not 
change (\ref{Amu}) :
\bea
 \label{Uk}
(B_1, B_2 ,I , J ) \longmapsto
(u B_1 u^{-1} , u B_2  u^{-1} , u I , J u^{-1} ) , 
 \qquad u \in U(k) .
\ena
Therefore
the moduli space of anti-self-dual
$U(n)$ gauge field  
with instanton number $k$ is given by
\bea
  \label{moduli}
{\cal M}_0 (k,n)
 = 
\mu^{-1}_{\R}(0) \cap \mu^{-1}_{\C}(0) /U(k) ,
\ena
where the action of $U(k)$ is the one
given in (\ref{Uk}). 
The fixed points of $U(k)$ action in
$\mu^{-1}_{\R}(0) \cap \mu^{-1}_{\C}(0)$
become singularities after the $U(k)$ quotients.
These singularities correspond to the instantons
shrinking to zero size,
and often called small instanton singularities
in physical literatures.

Let us check that the field strength
constructed from 
(\ref{Amu}) is really anti-self-dual:
\bea
F&=&
dA + A^2 \nn
&=&
d (\Psi^{\dagger} d \Psi )
+ (\Psi^{\dagger} d \Psi)( \Psi^{\dagger} d \Psi) 
\label{adsf1}\nn
&=& 
d \Psi^{\dagger} (1- \Psi\Psi^{\dagger})d \Psi.\label{adsf2}
\ena
In the above we have suppressed the
$U(n)$ indices.
One of the important points
in the ADHM construction
is that
$(1- \Psi\Psi^{\dagger})$ is a projection
acting on $V\oplus V\oplus W \approx 
\C^{2k + n}$ and project out 
the space of zero-modes ($\approx \C^n$).
Hence it can be rewritten as
\bea
 \label{projADHM}
1- \Psi\Psi^{\dagger} &=& 
{\cal D}_z^{\dagger}
\frac{1}{ {\cal D}_z {\cal D}_z^{\dagger}  }
{\cal D}_z  \nn
&=&
\tau^{\dagger}_z \frac{1}{\, \tau_z\tau^{\dagger}_z} \tau_z
+
\sigma_z \frac{1}{\, \sigma_z^{\dagger} \sigma_z}
\sigma_z^{\dagger } \nn
&=&
\tau^{\dagger}_z \frac{1}{\, \, \Box_z} \tau_z
+
\sigma_z \frac{1}{\, \, \Box_z}
\sigma_z^{\dagger },
\ena
where we have used the notations in (\ref{key}).
Since $\tau_z \Psi = \sigma_z^{\dagger } \Psi =0$
by definition (\ref{czerom}), it follows that
$\tau_z  d \Psi = -d \tau_z \Psi,\, 
 \sigma_z^{\dagger} d \Psi 
= - d \sigma_z^{\dagger} \Psi$.
Hence
\bea
\label{cFS}
F  
&=&
d \Psi^{\dagger} (1- \Psi\Psi^{\dagger})d \Psi \nn
&=&
d \Psi^{\dagger}
\left( \tau^{\dagger}_z \frac{1}{\, \, \Box_z} \tau_z
+
\sigma_z \frac{1}{\, \, \Box_z}
\sigma_z^{\dagger } \right)
d \Psi \nn
&=&
 \Psi^{\dagger}
\left( d\tau^{\dagger}_z \frac{1}{\, \, \Box_z} d\tau_z
+
d\sigma_z \frac{1}{\, \, \Box_z}
d\sigma_z^{\dagger } \right)
\Psi \nn
&=&
\Psi^{\dagger }
\left(
 \begin{array}{ccc}
dz_1 \frac{1}{\, \, \Box_z}d\bar{z}_1 
+ d\bar{z}_2 \frac{1}{\, \, \Box_z} dz_2   & 
  -dz_1 \frac{1}{\, \, \Box_z} d\bar{z}_2 
       + d\bar{z}_2\frac{1}{\, \, \Box_z} 
dz_1 & 0 \\
-dz_2 \frac{1}{\, \, \Box_z} d\bar{z}_1 
+ d\bar{z}_1\frac{1}{\, \, \Box_z} dz_2 & 
  dz_2 \frac{1}{\, \, \Box_z} d\bar{z}_2 
  + d\bar{z}_1 \frac{1}{\, \, \Box_z} dz_1 & 0 \\
 0 & 0 & 0
 \end{array}
\right) 
\Psi \nn
&:=& F^{-}_{\mbox{\tiny ADHM}} .
\ena
$F^{-}_{\mbox{\tiny ADHM}}$ is anti-self-dual:
$F^{-}_{\mbox{\tiny ADHM}} + * F^{-}_{\mbox{\tiny ADHM}} = 0$.

Now let us construct $U(2)$ one-instanton solution
by the ADHM method.
A solution to the ADHM equations is given by
\bea
 \label{com21}
B_1=B_2=0,\quad  
I = (\rho \, \, \, 0) ,\, \,  J^{\dagger} = (0 \, \, \,\rho).
\ena
Then the Dirac-like operator ${\cal D}_z $ becomes
\bea
{\cal D}_z 
=
\left(
\begin{array}{cccc}
-z_2 & -z_1 &\rho & 0 \\
 \bar{z}_1 & -\bar{z}_2 & 0 & \rho
\end{array}
\right).
\ena
We can find following zero-mode:
\bea
\Psi_{\mbox{\tiny BPST}} = 
\left(
\begin{array}{cc}
\rho &0 \\
 0 & \rho \\
z_2 & z_1 \\
-\bar{z}_1 & \bar{z}_2
\end{array}
\right)\frac{1}{\sqrt{r^2 + \rho^2}}. \label{CpsiB}
\ena
The gauge field constructed from this zero-mode 
is nothing but the well known
BPST instanton \cite{BPST}:
\bea
A_{\mu \mbox{\tiny BPST}} &=& \Psi^{\dagger}_{\mbox{\tiny BPST}} 
\,\pa_{\mu} \Psi_{\mbox{\tiny BPST}} \nn
&=&   \frac{r^2}{r^2 + \rho^2}\,  g^{-1} \pa_{\mu} g ,
\ena
where $ r = \sqrt{ x^{\mu}x_{\mu} }$ and
\begin{align}
 \label{gsing}
g(x) = \frac{x^{\mu}\sigma_{\mu}} {r} 
= \frac{1}{r} \left(
 \begin{array}{cc}
 z_2 & z_1 \\
 -\bz_1 & \bz_2 
 \end{array}
\right) , \quad \,
\sigma_{\mu} = (i \tau_1 , i \tau_2 , i\tau_3, 1).
\end{align}
Here $\tau_i\, (i = 1,2,3)$ are Pauli matrices. 
The instanton number is classified
by winding number $\pi^{3}({U(2)})$:
\begin{align}
 \label{wind}
\frac{1}{16 \pi^2} 
\int d^4x\, \mbox{tr}\, F_{\mu\nu}\tilde{F}^{\mu\nu} 
&=
\frac{1}{16 \pi^2} \int d^4x\,  \pa_{\mu}  K^{\mu}  \nn
&=
-\frac{1}{24 \pi^2}
\int_{S^3}\,  \mbox{tr}\, 
 g^{-1}dg   g^{-1}d  g  g^{-1}dg  \nn
&= -1 .
\end{align}
Here $\tilde{F}_{\mu\nu} 
= \frac{1}{2}{\epsilon_{\mu\nu} }^{\rho\sigma}
F_{\rho\sigma}$ and
\bea
K^{\mu} = 
2\, \mbox{tr}\,
\epsilon^{\mu\nu\rho\sigma}
\left(
A_{\nu} \pa_{\rho} A_{\sigma} 
+ \frac{2}{3} A_{\nu} A_{\rho} A_{\sigma}
\right).
\ena
For later purpose 
let us consider the following
zero-mode which is not well defined 
at the origin $r =  0$:
\bea
\Psi_{sing} = 
\left(
\begin{array}{cc}
\rho \bar{z}_2 & - \rho z_1 \\
\rho \bar{z}_1 & \rho z_2\\
\rho & 0 \\
 0 & \rho
\end{array}
\right)\frac{1}{ r \sqrt{r^2 + \rho^2}}. \label{CpsiS}
\ena
$\Psi_{sing}$ and $\Psi_{\mbox{\tiny BPST}}$ are
related by the ``singular'' gauge transformation
\bea
 \label{singtr}
\Psi_{sing} = \Psi_{\mbox{\tiny BPST}} \, g^{-1}(x).
\ena
Note that this transformation  
is not continuous at the origin.
Therefore 
$\Psi_{sing}$ is not an appropriate 
zero-mode for the ADHM construction.
However in the next subsection
we will observe
that in the noncommutative case,
we can construct a zero-mode 
similar to $\Psi_{sing}$,
but well defined everywhere !
 
The gauge field constructed from $\Psi_{sing}$
is given by
\bea
A_{\mu \, sing} = gA_{\mu \mbox{\tiny BPST}}\, g^{-1} 
+ g g^{-1} \pa_{\mu}g g^{-1}
=(\frac{r^2}{r^2 + \rho^2} - 1 ) g \pa_{\mu} g^{-1}
= \frac{\rho^2}{r^2 + \rho^2} \, g \pa_{\mu} g^{-1} ,
\ena
which is singular at the origin.
Note that the winding of $A_{\mu \mbox{\tiny BPST}}$
is resolved by the singular
gauge transformation $g$.

\subsection*{$U(2)$ One-instanton Solution on Noncommutative $\R^4$ %
and 
MvN Gauge Transformation}

As we have seen in the previous subsection,
the moduli space of instantons 
${\cal M}_0 (k,n)$ in (\ref{moduli})
has small instanton singularities.
The resolution of these singularities
is given in \cite{iNakaj}.
The fixed points of $U(k)$ action are
removed when we add a constant
to the right hand side of (\ref{ADHM}):
\bea
 \label{ADHMzeta}
\mu_{\R} = \zeta , \, \, \quad \mu_{\C} = 0  .
\ena  
Then the quotient space
\bea
 \label{Mzeta}
{\cal M}_{\zeta } (k,n) =
\mu_{\R}^{-1}(\zeta \, \mbox{Id}_V ) \,   
\, \cap \mu_{\C}^{-1} (0) \, / U(k) . 
\ena
is no longer singular.
The modification in (\ref{ADHMzeta}) 
modifies the key equation (\ref{key})
if we use ordinary commutative coordinates on $\R^4$.
However it was found in \cite{NS}
that if we use noncommutative coordinates
$z_i, \bz_i \,  (i = 1,2)$ which satisfies
$[z_1,\bar{z}_1]+[z_2,\bar{z}_2]
= -\zeta$,
$\tau_z$ and $\sigma_z$
do satisfy (\ref{key}).

We define operator
${\cal D}_z :
(V \oplus V \oplus W ) \otimes {\cal A}
\rightarrow ( V \oplus V ) \otimes {\cal A}$ 
by the same formula (\ref{Dz}):
\bea
& &{\cal D}_z = 
\left(
 \begin{array}{c}
   \tau_z \\
   \sigma_z^{\dagger }
 \end{array}
\right) , \nn
& & \tau_z =
(\, B_2 - z_2 ,\, B_1 - z_1 , \, I \, ), \nn
& & \sigma_z^{\dagger}  =
( \, - (B_1^{\dagger}-\bar{z}_1) , \,
  B_2^{\dagger} - \bar{z}_2  , \, J^{\dagger} \, ) .
\ena
The operator 
${\cal D}_z {\cal D}^{\dagger }_z :
(V \oplus V) \otimes {\cal A}
\rightarrow
(V \oplus V) \otimes {\cal A}$
has a block diagonal form,
\bea
 \label{box}
{\cal D}_z {\cal D}_z^{\dagger } 
=
\left(
 \begin{array}{cc}
  \Box_z & 0   \\
     0   & \Box_z 
  \end{array} 
 \right), \qquad
\Box_z \equiv \tau_z \tau_z^{\dagger}
       = \sigma_z^{\dagger } \sigma_z  ,
\ena
which is a consequence of (\ref{key})
and important for the ADHM construction.
Next we look for solutions of the 
equation 
\bea
 \label{zeroPsi}
{\cal D}_z \Psi^{(a)} = 0  \quad
( a = 1, \ldots , n ),
\ena
where 
the components of
$\Psi^{(a)} $ are {\em operators}: \ %
$\Psi^{(a)} : 
{\cal A} 
\rightarrow (V \oplus V \oplus W )\otimes {\cal A}$.
There is an important property that 
$\Psi$ must satisfy (see (\ref{projADHM})):
\bea
 \label{imp}
1 -  \Psi \Psi^{\dagger} 
= {\cal D}_z^{\dagger}
\frac{1}{ {\cal D}_z {\cal D}_z^{\dagger}  }
{\cal D}_z  .
\ena
This equation 
contains following 
two requirements.
First,
$\Psi$ must contain all the
vector zero-modes \cite{mine} on the left: 
The {\bf vector zero-mode}
$\left| {\cal U} \right\ran$ 
is an element of
${\cal H}^{\oplus k}  \oplus
    {\cal H}^{\oplus k}  \oplus  
    {\cal H}^{\oplus n}$
which satisfies
\bea 
{\cal D}_z \left| {\cal U} \right\ran = 0.
\ena
The operator zero-mode
$\Psi$ can be constructed from vector zero-modes.
In the case when the gauge group is $U(1)$,
the vector zero-modes are fully classified
\cite{Nakaj}\cite{LecNakaj}.
Second, ({\ref{imp}) imposes
normalization condition 
for $\Psi$.
The feature peculiar to the noncommutative 
case is that there may be
some states in ${\cal H}$
which are annihilated by
$\Psi^{(a)}$ for some $a$ \cite{NS}\cite{mine}.
More precisely,
all the components of $\Psi^{(a)}$
annihilate those states.
Then we can normalize $\Psi^{(a)}$
only in the subspace of ${\cal H}$  which is not
annihilated by $\Psi^{(a)}$.
It means that $\Psi$ is normalized as
\bea
\label{ncond}
\Psi^{\dagger}\Psi = p ,
\ena
where $p \in \M_n(\A)$ is a projection to the
states which are not annihilated by $\Psi$.
If the operator zero-mode satisfies
(\ref{imp}) and normalized as 
(\ref{ncond}), we can construct anti-self-dual
gauge field $A_p$ by the formula
\bea
A_p = \Psi^{\dagger} (d\Psi) \Psi^{\dagger}\Psi.
\ena
$A_p$ is anti-self-dual as a gauge connection
for $p \A^n$, i.e. if we consider
the covariant derivative for $p \A^n$:
$D_p = pd + A_p$.
When the gauge group is $U(1)$, there
is a natural choice for the normalization condition:
\bea
 \label{normI}
\Psi^{\dagger}\Psi = p_{\cal I} , 
\ena
where $p_{\cal I}$ is a projection to the
ideal states described in \cite{mine}.
We call the zero-mode
{\bf normalized minimal operator zero-mode}
when it is normalized as in (\ref{normI}).
Then the covariant derivative for $p_{\cal I} \A$:
\bea
D_{p_{\cal I}} = p_{\cal I} d + A_{p_{\cal I}} 
\ena
gives anti-self-dual field strength.

Because of the associativity of the
operator multiplication,
there is a freedom for the choice of the operator zero-mode:
\bea
\Psi \rightarrow \Psi U^{\dagger} , 
\ena
where 
$U^{\dagger}U = p$, $UU^{\dagger} = q$ and 
$q \in \M_n(\A)$ is a projection.
It is apparent that
$\Psi U^{\dagger}$ satisfies (\ref{imp}) if
$\Psi$ satisfies (\ref{imp}):
\begin{align}
\Psi U^{\dagger}  (\Psi U )^{\dagger}  
= \Psi U^{\dagger}U \Psi^{\dagger}  
= \Psi p \Psi^{\dagger}             
= \Psi \Psi^{\dagger} .
\end{align}
It is also easily seen that
this change of zero-modes corresponds
to the MvN gauge transformation.
Indeed,
\begin{align}
A_p = \Psi^{\dagger}(d\Psi) (\Psi^{\dagger}\Psi)
\rightarrow &\, 
(U\Psi^{\dagger}) (d (\Psi U^{\dagger}) ) 
(U\Psi^{\dagger}\Psi U^{\dagger}) \nn
&=
U(\Psi^{\dagger}d \Psi )U^{\dagger} ( U \Psi^{\dagger}\Psi U^{\dagger} )
+
U \Psi^{\dagger}\Psi (dU^{\dagger}) (U \Psi^{\dagger}\Psi U^{\dagger}) \nn
&=
U A_p U^{\dagger} + U (dU^{\dagger}) q,
\end{align}
which is nothing but the
MvN gauge transformation (\ref{Aq}).

Although we can choose arbitrary ``MvN gauge'' 
(or arbitrary projection),
there are not so many gauge choices which are
convenient or physically interesting.
In the case of $U(1)$ instanton,
the most natural choice may be
the one that corresponds to the
projection to the ideal states.
However, in the case of $U(2)$ instanton,
there is another choice which is 
physically interesting,
as will be  explained below.

Let us construct $U(2)$ one-instanton solution
by the ADHM method.
From hereafter we set $\zeta = 2$.
A solution to the 
ADHM equation (\ref{ADHMzeta}) is given by
\bea
B_1 = B_2 = 0, \quad
I = (\sqrt{\rho^2 + 2} \, \, \, \, 0\, ), \, \,
J = (\, 0\, \, \,  \rho  \,)  .
\ena
Then the Dirac-like operator ${\cal D}_z$
becomes
\bea
{\cal D}_z 
=
\left(
\begin{array}{cccc}
-z_2 & -z_1 & \sqrt{\rho^2 + 2} & 0 \\
 \bar{z}_1 & -\bar{z}_2 & 0 & \rho
\end{array}
\right).
\ena
The operator zero-mode can be obtained as
\begin{align}
\Psi_{min} = 
\left(
 \begin{array}{cc}
 & \\
\Psi^{(1)}_{min} & \Psi^{(2)}_{min}\\
 &
 \end{array}
\right)      \, ,       
\quad &\Psi^{(1)}_{min}
=
\left(
\begin{array}{c}
\sqrt{\rho^2 + 2}\, \bz_2 \\
\sqrt{\rho^2 + 2}\,  \bz_1   \\
z_1\bz_1 + z_2 \bz_2  \\
 0
\end{array} 
\right) \frac{1}
{ \sqrt{\hN (\hat{N}+2 + \rho^2)}  }, \nn
&\Psi^{(2)}_{min} = 
\left(
\begin{array}{c}
 -\rho z_1 \\
  \rho z_2 \\
    0      \\
z_1\bz_1 + z_2 \bz_2 +2
\end{array}
\right)  \frac{1}
{ \sqrt{ (\hN+ 2) (\hat{N}+2 + \rho^2)}  } ,
\label{SDpsiS}
\end{align}
where $\frac{1}{\sqrt{\hN} }$ is 
defined as
\bea
 \label{invN}
\frac{1}{\sqrt{\hN} }
=
\sum_{(n_1,n_2) \ne (0,0)}
\frac{1}{\sqrt{n_1 + n_2}} \,
| n_1,n_2 \ran\lan n_1,n_2| ,
\ena
i.e. when we consider the 
inverse of $\sqrt{\hN}$
we omit 
the kernel of $\sqrt{\hN}$, that is,
$|0,0\ran$, 
from the
Hilbert space. 
Hence $\frac{1}{\sqrt{\hN} }$ is a well defined operator.
This is an essential point in the
construction of instantons on noncommutative $\R^4$ \cite{mine}.
When $\rho = 0$, the contribution of $\Psi_{min}^{(2)}$ 
to the field strength vanishes whereas
$\Psi_{min}^{(1)}$ reduces to the normalized
minimal operator zero-mode in $U(1)$ one-instanton 
solution \cite{mine}.
The operator zero-mode $\Psi_{min}$
is normalized as 
\bea
\Psi^{\dagger}_{min} \Psi_{min} = p, 
  \label{NpsiS}
\ena
where $p$ is a projection in $\M_2(\A)$:
\bea
p = 
\left(
\begin{array}{cc}
\mbox{Id}_{\cal H}- | 0,0 \ran\lan 0,0 | & 0 \\
 0 & \mbox{Id}_{\cal H}
\end{array}
\right).
\ena
Although in the case where the gauge group
is $U(2)$ the vector zero-modes have not been
classified at the moment,
we can directly check that the equation (\ref{imp})
holds in this case:
\bea
{\cal D}_z^{\dagger}
\frac{1}{ {\cal D}_z {\cal D}_z^{\dagger}  }
{\cal D}_z
=
1 - \Psi_{min} \Psi^{\dagger}_{min}  .
\ena
Therefore the connection 
\begin{align}
D_p &= pd + A_p, \nn
&A_p = \Psi_{min}^{\dagger} (d \Psi_{min} ) 
 (\Psi_{min}^{\dagger} \Psi_{min})
\end{align}
gives anti-self-dual field strength.

Since the projection $p$ has infinite rank
as an operator in $\M_2(\A)$, it is
Murray-von Neumann equivalent to the
identity operator $\mbox{Id}_{\M_2(\A)}$.
So let us MvN gauge transform
$p$ to $\mbox{Id}_{\M_2(\A)}$.
In order to do so 
one seeks for the operator $U \in \M_2(\A)$ which
satisfies
\bea
 \label{UUp}
U^{\dagger}U = p, \quad  UU^{\dagger} = \mbox{Id}_{\M_2(\A)}.
\ena
Of course there are (infinitely) many choices for such $U$.
However there is a choice which has a physically interesting
interpretation.
Let us consider following operator $U$ 
which satisfies (\ref{UUp}):
\bea
U^{\dagger} = 
\left(
 \begin{array}{cc}
\frac{1}{\sqrt{\hN} } z_2 & \frac{1}{\sqrt{\hN} } z_1 \\
-\frac{1}{\sqrt{\hN+2 } } \bz_1 & \frac{1}{\sqrt{\hN +2}  } \bz_2
 \end{array}
\right).
\ena
Notice the similarity between this operator
and (the inverse of) the singular gauge transformation
(\ref{singtr}) in the commutative case.
However the operator $U$ is well defined unlike (\ref{gsing})
(remember that $\frac{1}{ \sqrt{\hat{N}} }$ is defined 
 as in (\ref{invN})).
Hence the MvN gauge transformation in this case
can be understood as a noncommutative resolution
of the singular gauge transformation (\ref{singtr}) !

After the MvN gauge transformation
the covariant derivative
takes the familiar form, i.e.
without projection operator on the left side of the derivative:
\bea
D = d + A.
\ena
Here the gauge field $A$ is constructed from the
gauge transformed zero-mode 
$\Psi_{\mbox{\tiny BPST}^*} = \Psi_{min} U^{\dagger}$:
\bea
A = A_{\mbox{\tiny BPST}^*} 
= \Psi_{\mbox{\tiny BPST}^*}^{\dagger} d \Psi_{\mbox{\tiny BPST}^*} .
\ena
If we express the gauge fields using operator symbols,
the long $r$ behavior of
$A_{\mbox{\tiny BPST}^*}$ is the same as that of the
BPST instanton
$A_{\mbox{\tiny BPST}}$ in commutative case,
and the instanton number 
$ \frac{1}{8 \pi^2}
\int d^4x\,
\mbox{tr}\,
F \wedge F$
is classified by $\pi^3(U(2))$, as in (\ref{wind}).
On the other hand the large $r$ behavior of
$A_p$ which is constructed from
$\Psi_{min}$
is the same as the one in
singular configuration $A_{sing}$ in commutative case.
Therefore the instanton 
number is not classified by $\pi^3(U(2))$ in this gauge.
However the instanton number itself does not change
under the MvN gauge transformation,
and in this case the instanton number count the dimension of 
the projection $(1-p)$, as described below.

We define new gauge field ${A'}_{\mu}$
for a notational convenience:
\bea
p(\hat{\pa}_{\mu}+ A_{\mu\, p})p
&=&
\hat{\pa}_{\mu}
-(1-p)\hat{\pa}_{\mu}-(1-p)\hat{\pa}_{\mu}
+(1-p)\hat{\pa}_{\mu}(1-p) +A_{\mu \, p}      \nn
&=&
\hat{\pa}_{\mu} + {A'}_{\mu}.
\ena
Here $\hat{\pa}_{\mu}$ is the derivative
operator (\ref{deri}). 
${A'}$ is {\it not} an MvN gauge transform of $A_p$.
The field strength of ${A'}_{\mu}$ is given as
\bea
{F'}_{\mu\nu}
&=&
[\hat{\pa}_{\mu} + {A'}_{\mu},\hat{\pa}_{\nu} + {A'}_{\nu}]
-[\hat{\pa}_{\mu},\hat{\pa}_{\nu}]  \nn
&=&
p ( iB_{\mu\nu} + F_{\mu\nu \, min}^{-}) p - i B_{\mu\nu} \nn
&=&
(1-p)iB_{\mu\nu}(1-p) + F_{\mu\nu \, min}^{-}  ,
\ena
where $F_{\, min}^{-} = D_p^2$.
It can be shown that the operator symbol of 
$A_{p}$ 
decays like $O(r^{-3})$ for large $r$
(recall the resemblance
between the minimal zero-mode (\ref{SDpsiS})
and the singular zero-mode (\ref{CpsiS})).
The operator symbol of 
$(1-p)$ decays like $\sim e^{-\frac{2}{\zeta}r^2}$.
$\theta^{\mu\nu}$
in the star product
appear as a multiplication of the combination 
$\theta/r^2$ for large $r$
and hence does not contribute to the 
surface integral at large $r$.
Taking all these accounts,
the instanton number of $A'$ vanishes:
\bea
\frac{1}{16 \pi^2}
\left(2\pi\right)^2 \sqrt{ det \theta}\,
\mbox{Tr}\,
{F'}_{\mu\nu} {\tilde{F}}^{'\mu\nu}
&=&
\frac{1}{16 \pi^2}
\int d^4x\,
\mbox{tr}\,
{F'}_{\mu\nu} \star {\tilde{F}}^{'\mu\nu} \nn
&=&
\frac{1}{16 \pi^2}
\int d^4x\,
\pa_{\mu} K^{\mu} = 0.
\ena
Here 
\bea
K^{\mu} = 
2\, \mbox{tr}\,
\epsilon^{\mu\nu\rho\sigma}
\left(
A'_{\nu}\star \pa_{\rho} A'_{\sigma} 
+ \frac{2}{3} A'_{\nu}\star A'_{\rho} \star A'_{\sigma}
\right).
\ena
On the other hand,
\begin{align}
&\, \frac{1}{16 \pi^2}
\left(2\pi\right)^2 \sqrt{ det \theta}\, \,
\mbox{Tr}\, \,
{F'}_{\mu\nu} {\tilde{F}}^{'\mu\nu} \nn
=&\, 
\frac{1}{16 \pi^2}\left(2\pi\right)^2
\sqrt{det \theta}\, \, \mbox{Tr}\,
\left[
(1-p)B_{\mu\nu}\tilde{B}^{\mu\nu}(1-p) -
 F_{\mu\nu \, min}^- \tilde{F}^{\mu\nu -}_{\, min}
\right],
\end{align}
($B_{\mu\nu}$ is self-dual
as we have set $\theta^{\mu\nu}$ self-dual).
Thus the instanton number counts the
dimension of the projection $(1-p)$:
\begin{align}
&\, \frac{1}{16 \pi^2}
 \left(2\pi\right)^2 \sqrt{ det \theta}\,
 \mbox{Tr}\,
F_{\mu\nu \, min}^- F^{\mu\nu -}_{\, min}  \nn
=&\, 
\frac{1}{16 \pi^2}
\left(2\pi\right)^2 \sqrt{ det \theta}\,
\mbox{Tr}\,
(1-p)B_{\mu\nu}\tilde{B}^{\mu\nu}(1-p) \nn
=&\,  \mbox{dim}\, (1-p).
\end{align}

\section{Conclusion}

In this article the formalism that 
describes the equivalence of projections 
as a kind of gauge equivalence on noncommutative 
space is given. 
We apply this formalism to the
$U(2)$ one-instanton solution
on noncommutative $\R^4$.
The gauge equivalence between
BPST type configuration 
with winding number one and 
the configuration without winding
but with projection is shown.
In this case the gauge transformation
can be understood as a noncommutative
resolution of the singular gauge transformation
in ordinary $\R^4$.
Recall that the projection 
describes  holes on 
noncommutative $\R^4$ \cite{mine}.
Hence this formalism gives 
a unified description to the 
intriguing mixing of gauge fields
and geometry in noncommutative space 
\cite{mine}\cite{BN}.

\vs{5}

\begin{center}
{\bf \large Acknowledgments}
\end{center}
I would like to thank N. Ishibashi and S. Iso
for useful discussions. 
I would also like to thank
T. Hirayama and Y. Okada
for suggestions and encouragements.


\newcommand{\NP}[1]{Nucl.\ Phys.\ {\bf #1}}
\newcommand{\AP}[1]{Ann.\ Phys.\ {\bf #1}}
\newcommand{\PL}[1]{Phys.\ Lett.\ {\bf #1}}
\newcommand{\CQG}[1]{Class. Quant. Gravity {\bf #1}}
\newcommand{\CMP}[1]{Comm.\ Math.\ Phys.\ {\bf #1}}
\newcommand{\PR}[1]{Phys.\ Rev.\ {\bf #1}}
\newcommand{\PRL}[1]{Phys.\ Rev.\ Lett.\ {\bf #1}}
\newcommand{\PRE}[1]{Phys.\ Rep.\ {\bf #1}}
\newcommand{\PTP}[1]{Prog.\ Theor.\ Phys.\ {\bf #1}}
\newcommand{\PTPS}[1]{Prog.\ Theor.\ Phys.\ Suppl.\ {\bf #1}}
\newcommand{\MPL}[1]{Mod.\ Phys.\ Lett.\ {\bf #1}}
\newcommand{\IJMP}[1]{Int.\ Jour.\ Mod.\ Phys.\ {\bf #1}}
\newcommand{\JHEP}[1]{J.\ High\ Energy\ Phys.\ {\bf #1}}
\newcommand{\JP}[1]{Jour.\ Phys.\ {\bf #1}}

\end{document}